\documentstyle[12pt,cite,epsf,epsfig]{article}
\newcommand{\newc}{\newcommand}
\newc{\tev}{\,{\rm TeV}}
\newc{\gev}{\,{\rm GeV}}
\newc{\sgn}{\mr{sgn}\,}
\newc{\ra}{\rightarrow}
\newc{\rpv}{$\mathrm{\not\!R_p}$}
\newc{\met}{$\not\!\!E_T$}
\newc{\rp}{$\mathrm{R_p}$}
\newc{\real}{\mathcal{R}e}
\newc{\alsm}{{\displaystyle \sum_{\alpha=1,2}}}
\newc{\besm}{{\displaystyle \sum_{\beta=1,2}}}
\newc{\al}{\alpha}
\newc{\ga}{\gamma}
\newc{\de}{\delta}
\newc{\cw}{\cos\theta_w}
\newc{\ssw}{\sin^2\theta_w}
\newc{\ccw}{\cos^2\theta_w}
\newc{\cbe}{\cos\beta}
\newc{\sbe}{\sin\beta}
\newc{\sh}{\hat{s}}
\newc{\sa}{\sin\al}
\newc{\ca}{\cos\al}
\newc{\bv}{$\mathrm{\not\!B}$}
\newc{\lv}{$\mathrm{\not\!L}$}
\newc{\ie}{{\it i.e.\/}\ }
\newc{\lam}{\lambda}
\newc{\cht}{\tilde{\chi}}
\newc{\upt}{\tilde{u}}
\newc{\elt}{\tilde{\ell}}
\newc{\hgt}{\tilde{H}}
\newc{\nut}{\tilde{\nu}}
\newc{\dnt}{\tilde{d}}
\newc{\psb}{\bar{\psi}}
\newc{\rtt}{\sqrt{2}}
\newc{\mut}{\tilde{\mu}}
\newc{\mr}{\mathrm}
\newc{\bath}{\bar{\theta}}
\newc{\tht}{\theta}
\newc{\JC}{{\bf J}}
\newc{\lra}{\longrightarrow}
\newc{\eg}{{\it e.g.\,}}
\newc{\barr}{\begin{array}}
\newc{\earr}{\end{array}}
\newc{\dis}{\displaystyle}
\newc{\beq}{\begin{equation}}
\newc{\eeq}{\end{equation}}
\newc{\me}{\mathcal{M}}
\newc{\dbm}{\partial_\mu}
\newc{\sgm}{\sigma_\mu}

\def\ra{\rightarrow}
\def\dis{\displaystyle}
\catcode`@=11 % This allows us to modify PLAIN macros.
\def \gsim{\mathrel{\mathpalette\@versim>}}
\def \lsim{\mathrel{\mathpalette\@versim<}}
\def \@versim#1#2{\lower0.4ex\vbox{\baselineskip\z@skip\lineskip\z@skip
     \lineskiplimit\z@\ialign{$\m@th#1\hfil##\hfil$%
     \crcr#2\crcr\sim\crcr}}}
\catcode`@=12 % at signs are no longer letters
\def\gev{\: \rm GeV}
\begin{document}
\setcounter{page}{0}
\renewcommand{\thefootnote}{\fnsymbol{footnote}}
\thispagestyle{empty}

\begin{titlepage}
\vspace{-2cm}
\begin{flushright}
BU-HEPP-07-03\\[2ex]
%{\large \tt hep-ph/yymmnnn}\\
\end{flushright}
\vspace{+2cm}

\begin{center}
 {\Large{\bf Dijet production in generic contact interaction\\[1ex]
 at linear colliders}
}\\
\vskip 0.6 cm
{\bf %Debajyoti Choudhury$^{1,3}$ {\rm and} 
Swapan Majhi}
\footnote{E-mails: \ swapan\_majhi@baylor.edu}
        \vskip 1cm
%$^{1}${\it Department of Physics and Astrophysics, University of Delhi,
%       Delhi 110 007, India.}\\[1ex]

{\it Department of Physics, Baylor University, Waco, TX 76706, U.S.A.
          }
\\
\vskip 0.2 cm
\end{center}
\setcounter{footnote}{0}
\begin{abstract}\noindent
We consider dijet production at a $e^+ e^-$ collider in a class
of effective theories with the relevant operators being four-fermion
contact interaction. Despite the nonrenormalizable nature of the interaction,
we explicitly demonstrate that calculating QCD corrections is both possible
and meaningful. Calculating the corrections for various
differential distributions, we show that these can be substantial
and significantly different from those within the SM. Furthermore,
the corrections have a very distinctive flavor dependence.
\end{abstract}
\end{titlepage}
\setcounter{footnote}{0}
\renewcommand{\thefootnote}{\arabic{footnote}}

\setcounter{page}{1}
\pagestyle{plain}
\advance \parskip by 10pt

\section{Introduction}

Measurement of multi-jets rates in electron-positron annihilation provide an excellent  test of perturbative quantum chromodynamics (QCD)\cite{Kram_Lam_1_6}. In Standard Model (SM), QCD has been tested in the perturbative regime to a high degree of accuracy\cite{Bel_Cam_Ros_1}. 
Though SM is very successful model in high energy physics, there are theoretical
issues that cannot even be addressed within the framework of the SM
alone. Examples include the replication of the fermion families, the
naturalness problem associated with the Higgs scale, charge
quantization, the baryon asymmetry in the universe, the presence of
dark matter etc.. Clearly, an answer to such vital questions may be
obtained only in a model much more ambitious than the SM. Candidates
for the role include, amongst others, supersymmetry~\cite{susy}, grand
unification~\cite{Pati:1974yy,GUTS} (with or without supersymmetry),
family symmetries (gauged or otherwise) and compositeness for quarks
and leptons~\cite{llmodel}.
Therefore the possible existence of new physics beyond SM, involving the four 
fermion contact interaction would be one of the viable model and  may give 
rise to small effects at present and future linear collider. We 
illustrate this explicitly in the case of dijet production via 
electron-positron annihilation process. We limit our calculation 
for exclusive two-jet cross section in this model.

The replication of fermion families in SM suggests the possibility of
quark-lepton composite structure or  bound states of more fundamental
constituents, called {\em preon}s\cite{preon}.
't Hooft has shown that the gauge theory\cite{preon_binding} of preon binding naturally
reproduce the composite fermions of less massive than the preon binding scale
which is called characteristic scale $\Lambda$. At this
scale $\Lambda$, this interaction would become very strong leading to
bound states (composites) which are to be identified as quarks and
leptons.  In most such models\cite{comp_mod, Additional_comp}, quarks
and leptons share at least some common constituents.  Since the
confining force mediates interactions between such constituents.
% it
%stands to reason that these, in turn, would lead to interactions
%between quarks and leptons that go beyond those existing within the SM.
Well below the scale $\Lambda$, such interactions would likely be
manifested through an effective four fermion contact
interaction~\cite{cont_inter} term
that is an invariant under the SM gauge group.
A convenient and
general parametrization of such interactions is given
by\cite{llmodel,Taekoon}
\beq 
{\cal L} = {4 \pi \over \Lambda^2} \Big[
\eta_{ij} \, (\bar{q} \, \gamma^{\mu} \, P_i \, q) \,
          (\bar{l} \, \gamma_{\mu} \, P_j \, l)
+ \xi_{ i j }  \, (\bar q \, P_i \, q) \, (\bar \ell \, P_j  \, \ell)
 \Big],
     \label{lagrangian}
\eeq
where $i, j = L, R$ and $P_i$ are the chirality projection operators.
Note that the Lagrangian of eqn.(\ref{lagrangian}) is by no means a
comprehensive one and similar operators involving the quarks alone (or
the leptons alone) would also exist.
However, for our purpose, it
would suffice to consider only eqn.(\ref{lagrangian}). Within this
limited sphere of applicability, the strength of the interaction may
be entirely absorbed in the scale $\Lambda$, and the couplings
$\eta_{ij}$ and $\xi_{ij}$ canonically normalized to $\pm 1$.

Though the lagrangian presented above eqn.(\ref{lagrangian}) is  
so-called {\it 4-fermion} contact interaction lagrangian but there are other theories
which can give rise to such an effective interaction lagrangian. As is well
known, a four-fermion process mediated by a particle with a mass
significantly higher than the energy transfer can be well approximated
by a contact interaction~\cite{cont_inter} term with a generic form as
in eqn.(\ref{lagrangian}).
For examples, theories with extended
gauge sectors, leptoquarks~\cite{leptoquarks}, sfermion
exchange in a supersymmetric theory with broken
$R$-parity~\cite{fayet} etc. are the theories which can give rise to 
such kind of effective interaction lagrangian by integrating out fields with masses 
$M_i \gsim \Lambda$\cite{Hasenfratz:1987tk}. 
In those eqn.(\ref{lagrangian}) are just the lowest order (in $\Lambda^{-1}$) 
ones among the series of
such higher-dimensional terms.

%In all such cases, on integrating out
%fields with masses $M_i \gsim \Lambda$\cite{Hasenfratz:1987tk}, a
%series of such higher-dimensional terms obtain. Those in
%eqn.(\ref{lagrangian}) are just the lowest order (in $\Lambda^{-1}$)
%ones.

Such operators, in principle, could lead to
significant phenomenological consequences in collider experiments,
whether $e^+ e^-$~\cite{opal,opal2}, $e \, P$~\cite{HERA2} or hadronic.
Given the higher-dimensional nature of ${\cal L}$, 
%it is obvious that
%the consequent effects would be more pronounced at higher energies. 
%In otherwords, 
the fractional deviation over the SM expectations would 
be concentrated more at higher invariant masses. 
Some of the best constraints on compositeness,
for example, came from the CDF~\cite{cdfprl} experiments and more recent 
measurement of the Drell-Yan cross section\cite{Drell_Yan} at high
invariant masses set the most stringent limits \cite{D0_coll, NKMondal} on 
contact interactions
of the type given in eqn.(\ref{lagrangian}). For the  $e^+ e^-$ collider,
more recent constraints on compositeness, for example came from the 
OPAL \cite{opal2} $\Lambda \gsim 1.6$--3.4 TeV within
the $VA$-type interaction scenario.

Recently, we have done the NLO QCD correction \cite{3rd_paper} in the context of hadron collider by 
taking into account of this effective lagrangian eqn.(\ref{lagrangian}). In this article,
 we are going to do similar type of calculation in Linear Collider.
In $e^+ e^-$ annihilation, the perturbative QCD predicts only parton cross section but 
experimentally one measured only hadrons though hadronisation process known only 
phenomenologically. Because of this limited knowledge of hadronisation process one can 
not directly relate theory and experiment. Since we measured only hadrons in the final 
state one should include the higher order QCD corrections (which include the more 
partons in the final state) to the lowest order one to get the better result. In SM,
people have done their calculation of next-to-leading order 
(NLO)\cite{sterman,Giele_Glover} and next-to-next-to-leading 
order (NNLO)\cite{kramer_lampe} QCD corrections for the dijet production in 
$e^+ e^-$ annihilation. However, no calculations exist for the higher order 
QCD corrections to cross sections mediated by a  generic contact interaction.
Consequently, all
extant collider studies of contact interaction have either been based on
just the tree level calculations,  or, in some cases, assumed
that the higher order corrections are exactly the same as in the SM.
Clearly, this is an unsatisfactory state of affairs and, in this paper,
we aim to rectify this by calculating the next-to-leading order QCD
corrections for both the $VA$-type and the $SP$-type contact interactions.

%It might be argued that, such theories being nonrenormalizable, any
%higher-loop calculation is fraught with danger. However, the very structure
%of such terms (namely the current--current form of the Lagrangian) along with
%the fact that only one of the currents comprises coloured fields allows
%us to reliably calculate  QCD corrections. This holds not only for
%the specific interaction in question, but also for other theories that
%satisfy the abovementioned criterion~\cite{ravi_gravi}. 

Being nonrenormalisable of such theories, the current-current form of
the lagrangian allow us to calculate reliable QCD corrections because 
of the fact that only one
current consists of coloured field.This holds not only for
the specific interaction in question, but also for other theories that
satisfy the abovementioned criterion~\cite{ravi_gravi}. 

The rest of the article is organised as follows. In Section \ref{NLO_corr}, we
present our results for the LO and NLO cross section for resolved two parton case only (stated otherwise). The resolved three parton cross section will be divergence free and hence can be evaluated numerically in 4-dimension. Here we also present only the total three parton cross section.
In Section \ref{res_dis} we discuss the numerical results and 
finally, we summarise in Section \ref{summery}.

\section{NLO corrections}
\label{NLO_corr}
    Before going to actual calculation of jet cross section, it is necessary to define cross section in perturbation theory. In perturbative QCD, each outgoing hard parton regarded as one jet that means one has to apply a jet resolution criterion to each outgoing partons to define a jet. In other words, a proper definition  is to introduce a parton resolution criterion to define when a parton is resolved either as a single parton or as a cluster of partons. Consider the process $e^+e^-$ annihilate to quarks and gluon i.e. $e^+e^- \rightarrow q(p_{q}) \; \bar{q}(p_{\bar{q}})\; g(p_{g})$. This process can be thought of as lowest order three jet production or higher order dijet production depending upon how we define the jet resolution criterion. One possible jet definition is a minimum mass cut so that the invariant mass of pair of jet must be larger than experimentally defined one $s_{min}$. In the above mentioned process,
there are only three invariant mass $s_{ij} = (p_i + p_j)^2, i,j = q,\bar{q},g$. Therefore the lowest order three jet cross section is defined by
\beq
d\sigma (e^+e^- \rightarrow 3\; jets) = \Theta d\sigma (e^+e^- \rightarrow q \bar{q} g)
\eeq
and $\Theta$ is the jet resolution criterion for the three jet final state defined by
\beq
\Theta = \theta(s_{min}-s_{q\bar{q}})\;\theta(s_{min}-s_{qg})\;\theta(s_{min}-s_{\bar{q}g})
\eeq
where $\theta(x) = 1$ for $x > 0$ and 0 otherwise.
For the dijet cross section, it will represent the next-to-leading order (NLO) dijet cross section (where one of the partons (gluon, say) is either soft or collinear to other partons) and the jet resolution criterion for the two-jet final state would be the only one of the invariant masses ($s_{q\bar{q}}$ say) larger than the experimentally defining cut $s_{min}$ i.e.
\beq
d\sigma (e^+e^- \rightarrow 2\; jets) = \Theta d\sigma (e^+e^- \rightarrow q \bar{q} g)
\eeq
where 
\beq
\Theta = \theta(s_{min}-s_{q\bar{q}}).
\eeq

We consider the process $e^+e^-$ annihilation into a quark-antiquark pair i.e. $e^+ e^- \rightarrow q \,\bar{q}$ in the context of generic contact interaction as defined by the lagrangian eqn.(\ref{lagrangian}).
In the presence of scalar-pseudoscalar ($SP$) type contact interaction, the leading order differential cross section for the above process is given by
\beq
\barr{rclcl c rclcl}
{d \sigma^{(0)}_{SP} \over d \cos \theta} &=& {3 \pi \over 2 } 
\sum_{i,j=L,R}|\xi_{ij}|^2\;{s \over \Lambda^4}
\label{diff_lo_cs_sp}
\earr
\eeq
and for the vector-axial-vector ($VA$) type contact interaction, the leading order differential cross section for that same process (stated above) will be same as the standard model one and is given by (for completeness)
\beq
\barr{rclcl c rclcl}
{d \sigma^{(0)}_{VA} \over d \cos \theta} &=& {3 \pi \alpha^2 \over 2\,s }
\Bigg[ \Big(|f_{LL}|^2 + |f_{RR}|^2 \Big)\,\Bigg({u \over s}\Bigg)^2 +
\Big(|f_{LR}|^2 + |f_{RL}|^2 \Big)\,\Bigg({t \over s}\Bigg)^2 
\Bigg]
\label{diff_lo_cs_va}
\earr
\eeq
where $t = -{s \over 2} (1-\cos \theta)$, $u = -{s \over 2} (1+\cos \theta)$ and $\alpha$ is the electromagnetic coupling constant. $f_{ij} (i,j = L,R)$ are given by
\beq
\barr{rclcl c rclcl}
f_{ij} &=& Q_l Q_q + g_i^q g_j^l \chi(s) + \eta_{ij} {s \over \alpha \Lambda^2}.
\nonumber\\[2ex]
\chi(s) &=& s / ( s-M_Z^2 + i M_Z \Gamma_Z)
\earr
\eeq
The left-handed and right handed couplings $g_L^f$ and $g_R^f$ of the fermion to $Z$-boson are given by,
\beq
g_L^f = {e \over \sin\theta_W \cos\theta_W} \Big(I_3^f - Q_f \sin^2\theta_W \Big),
\hspace{0.5cm} 
g_R^f = {e \over \sin\theta_W \cos\theta_W} \Big(- Q_f \sin^2\theta_W \Big)
\eeq
where $e$ is the electron charge,$Q_f$ is the electric charge in units of $|e|$ of the fermion $f$, $I_3^f$ is the third component of weakisospin and $\theta_W$ is the electroweak mixing angle.

At the leading order (LO) dijet calculation is much more simpler (as calculated above eqns.(\ref{diff_lo_cs_sp},\ref{diff_lo_cs_va}) than the next-to-leading order. At NLO, it requires careful treatment of cancellation of divergences (soft and collinear divergences against the divergences stemming from virtual corrections). The divergences coming from the fact that at NLO a parton can only be defined through a resolution criterion. There are many form of this resolution criterion. We have used the invariant mass resolution criterion. That is, if the invariant mass of the two parton less than the invariant mass resolution ($s_{min}$) then these two parton considered as unresolved parton and treated as a one parton (or one jet) by integrating out the unresolved phase space which separates the soft and collinear region of phase space from the resolved bremsstrahlung phase space. After adding this unresolved soft and collinear contribution with virtual corrections, it becomes finite. 

For dijet production, the order $\alpha_s$ correction receive two contributions, one is from resolved two parton cross section (which is purely virtual contribution) and other one is the lowest order unresolved three parton cross section (calculated in the  soft and collinear limit) where one soft and/or collinear parton clusters with one hard parton to form one jet. These soft and/or collinear divergences can be isolated and it is easy to show that these divergences analytically cancel with  soft and collinear singularities coming from virtual correction of resolved two part process upto order $s_{min}$ (or $y_{min} = s_{min}/s_{ij}$) where virtual gluon in the loop becomes soft. The lowest order unresolved three parton cross section will be leading order cross section multiplied by some functions $F^{S+C}$ (a function of all the singularities (soft and collinear), $s_{min}$ (or $y_{min}$) and the factorisation scale ($\mu_F$)).  This $F^{S+C}$ is same for $VA$-type or $SP$-type theory and can be found out in the literature (for example see \cite{Giele_Glover}). The lowest order three parton cross section is given by,

\beq
\barr{rclcl c rclcl}
d \sigma^{(S+C)}_{\eta} &=&d \sigma^{(0)}_{\eta} \times F^{S+C}
\nonumber\\[3ex]
&=& d \sigma^{(0)}_{\eta} \times {\alpha_s C_F \over 2 \pi \Gamma(1-\epsilon)}
\Bigg[{2 \over \epsilon^2}\,\Bigg({4 \pi \mu^2_F \over s_{q\bar{q}}}\Bigg)^{\epsilon} 
- 2\ln^2\Bigg({s_{q\bar{q}} \over s_{min}}\Bigg) + 7 - {2\,\pi^2 \over 3}
\nonumber \\[2ex]
&&+{3 \over \epsilon} \Bigg({4 \pi \mu^2_F \over s_{min}}\Bigg)^{\epsilon}
+{\cal O}\Big(s_{min} / s_{q\bar{q}}\Big)
\Bigg]
\hspace*{3.5cm} (\eta = SP, VA)
\label{soft_col:expr}
\earr
\eeq
  
       The virtual corrections to the resolved two parton process is also available in the literature (for example they can be read from Refs.\cite{1st_paper} for $SP$-type and Refs.\cite{Giele_Glover,ravi_gravi} for $VA$ type theory). We follow the dimensional regularisation procedure to regulate all the divergences in $d= 4-2 \epsilon$ dimensions and we use the $\overline{MS}$ scheme to remove the ultra-violet divergence. For completeness, they are given by
\beq
\barr{rclcl c rclcl}
 d \sigma^{(0+V)}_{\eta}  &=& d \sigma^{(0)}_{\eta} 
\Bigg(1+ {\alpha_s C_F \over 4\, \pi} F^{(1)}_{\eta}
\Bigg)
\hspace*{2cm} (\eta = SP, VA)
\label{eq:virtual}
\earr
\eeq
where 
\begin{equation}
F^{(1)}|_{VA}={ 2 \over \Gamma\left (1 - \epsilon \right)}
~\Bigg({s_{q\bar{q}} \over 4 \pi\,\mu^2}\Bigg)^{-\epsilon }
\Bigg[ -{2 \over \epsilon^2} ~-~{3 \over \epsilon}~ 
-8 + \pi^2 \Bigg] \ ,
\label{F:sm}
\end{equation}

\begin{equation}
F^{(1)}|_{SP}={ 2 \over \Gamma\left (1 - \epsilon \right)}
~\Bigg({s_{q\bar{q}} \over 4 \pi\,\mu^2}\Bigg)^{-\epsilon }
\Bigg[ -{2 \over \epsilon^2} ~-~{3 \over \epsilon}~ 
-2 + \pi^2 \Bigg] \ .
\label{F:scalar}
\end{equation}

From eqns.(\ref{soft_col:expr},\ref{eq:virtual}), it is clear that the left over soft and/or collinear divergences (from the virtual corrections) cancel against the soft and collinear divergences coming from unresolved three parton process. Therefore next-to-leading order cross section for exclusive dijet production are given by 
\beq
\barr{rclcl c rclcl}
d \sigma^{(1)}_{VA} &=& d \sigma^{(0)}_{VA} \times\Bigg[1+ {\alpha_s C_F \over 2 \pi \Gamma(1-\epsilon)}
\Bigg({\pi^2 \over 3} -1 - 3 \ln\Big(y_{min}\Big)
- 2\ln^2\Big(y_{min}\Big)
\Bigg) + {\cal{O}} \Big(y_{min}\Big)
\Bigg]
\label{sig_2jet_va}
\earr
\eeq
and 
\beq
\barr{rclcl c rclcl}
d \sigma^{(1)}_{SP} &=& d \sigma^{(0)}_{SP} \times\Bigg[1+ {\alpha_s C_F \over 2 \pi \Gamma(1-\epsilon)}
\Bigg({\pi^2 \over 3} + 5 - 3 \ln\Big(y_{min}\Big)
- 2\ln^2\Big(y_{min}\Big)
\Bigg) + {\cal{O}} \Big(y_{min}\Big)
\Bigg]
\label{sig_2jet_sp}
\earr
\eeq
which are finite as $\epsilon \rightarrow 0$. Here we have considered $\mu= \mu_F$.
By integrating over resolved three particle phase space, one can get the 
${\cal {O}}(y_{min})$ correction of the above 
eqns.(\ref{sig_2jet_sp},\ref{sig_2jet_va}). The resolved three parton cross section for the $VA$-type contact interaction will be same as $SM$ which is available in literature (see for example Ref.\cite{g_kramer,kramer_lampe}) and is given by,
\beq
\barr{rclcl c rclcl}
\sigma_{3-jet}^{VA} &=& \sigma^{(0)}_{VA}\;{\alpha_s C_F \over 2 \pi \Gamma(1-\epsilon)}\;
\Bigg[-{\pi^2 \over 3} + {5 \over 2} + 3 \ln\Big(y_{min}\Big)
+ 2\ln^2\Big(y_{min}\Big)
 + {\cal{O}} \Big(y_{min}\Big) \Bigg].
\label{sig_3jet_va}
\earr
\eeq
For the $SP$-type contact interaction we have calculated the resolved three parton cross section as given below
\beq
\barr{rclcl c rclcl}
\sigma_{3-jet}^{SP} &=& \sigma^{(0)}_{SP}\;{\alpha_s C_F \over 2 \pi \Gamma(1-\epsilon)}\;
\Bigg[-{\pi^2 \over 3} + {7 \over 2} + 3 \ln\Big(y_{min}\Big)
+ 2\ln^2\Big(y_{min}\Big)
 + {\cal{O}} \Big(y_{min}\Big) \Bigg].
\label{sig_3jet_sp}
\earr
\eeq
From eqns.(\ref{sig_2jet_va},\ref{sig_3jet_va}) (or eqns.(\ref{sig_2jet_sp},\ref{sig_3jet_sp})), it is clear that the theoretical results for resolved two parton and three parton depend strongly on an arbitrary parameter $y_{min}$. Any physical observable should not depend on this arbitrary parameter. However for physical $2$-jet NLO cross section, both two parton and three parton cross section will contribute and hence it is independent of this arbitrariness. This also ensures the KLN (Kinoshita-Lee-Nauenberg) theorem that the fully inclusive $e^+e^-$ cross section is finite as quark mass goes to zero (i.e. free of mass singular).

\section{Results and Discussions}
\label{res_dis}
        In this section, we present numerical result for ILC.
We choose the contact interaction scale ($\Lambda$) to be 2 TeV and the center of mass energy to be $\sqrt{S} = 500$ GeV. As is well known, the higher order QCD correction reduces the uncertainties related to scale choice (the renormalisation scale ($\mu$) and the factorisation scale ($\mu_F$)). For the NLO jet calculation, the analytical result does not depend on any of these scale explicitly\cite{catani}. The scale dependence comes through the strong coupling constant $\alpha_s(\mu^2)$.  We have used the NLO $\alpha_s(\mu^2)$ for the NLO analysis and the scale we choose, both renormalisation and factorisation scale to be $\mu ~(\mu_F) = P_T$ (otherwise stated). We have also shown that both two-parton and three parton result strongly depends on the cut-off $y_{min}$. For very small values of $y_{min}$, two parton cross section become negative and three parton cross section become large positive (because of these terms $\ln (y_{min}),~ \ln^2(y_{min})$). For large enough $y_{min}$, both the parts produce the meaningful result. Our analytical result is valid for small $y_{min}$ region since we have neglected the term ${\cal O}({y_{min}})$ in the integration. Therefore $y_{min}$ should be much less than one ($y_{min} << 1$). For this reason, we choose $y_{min} = 0.01$ (detailed discussion can be found in the literature \cite{Giele_Glover,kramer_lampe}) for all the  differential distributions and the total cross section. 
Furthermore, in presenting our results,
we shall consider only one of the couplings $\eta_{AB}$
and $\xi_{AB}$ to be non-zero and of unit strength.

For the sake of convenience, we parametrize the cross section as
\begin{equation}
\begin{array}{rclcl}
\sigma & = & \sigma_{\rm SM} + \sigma_{\rm intf} + \sigma_{\eta^2}
       & \qquad & {\rm (for \ the \ VA \ case)}
\\[2ex]
\sigma & = & \sigma_{\rm SM}  + \sigma_{\xi^2}
       & \qquad & {\rm (for \ the \ SP \ case)}
\end{array}
\end{equation}
and similarly for the differential cross sections. This has the
advantage in that the total cross sections, for an arbitrary
value of $\Lambda$ can be easily reconstructed. 
We also take care of the
so-called ${\it{initial ~ state ~ radiation}}$(ISR) effect\cite{ISR} through out our numerical analysis
otherwise stated.

%%%%%%%%
\begin{figure}[!h]
\centerline{
\epsfxsize=16cm\epsfysize=16.0cm
                     \epsfbox{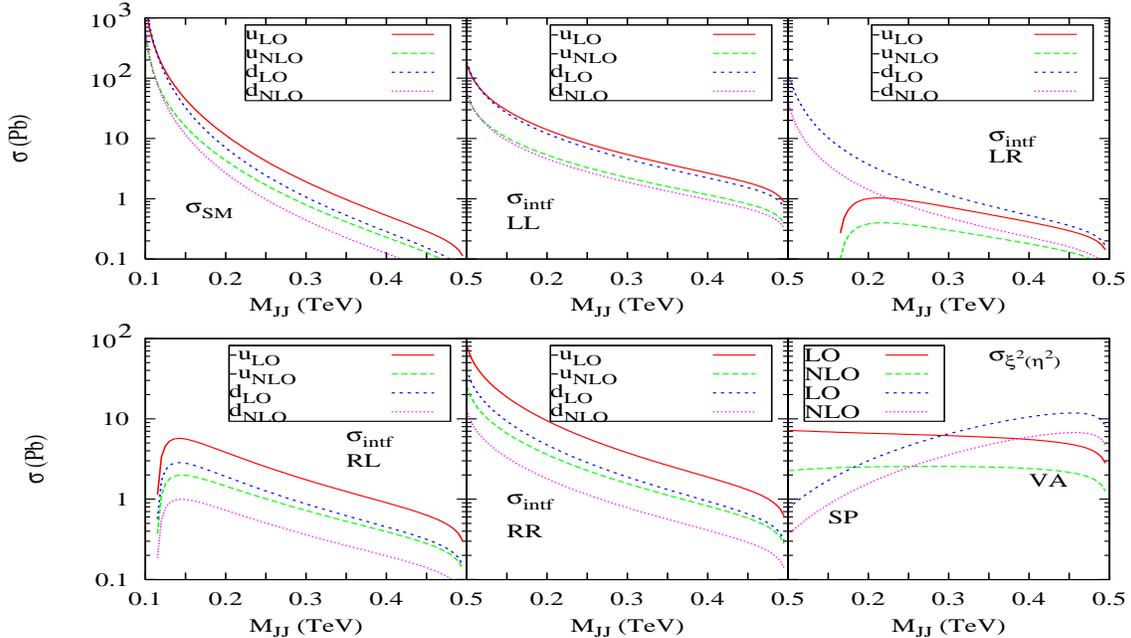}
}
\vspace*{-6.5cm}
\caption{\em The dijet production cross section as a function of invariant mass $M_{JJ}$ for $\sqrt{S} = 500$ GeV and $\Lambda = 2$ TeV. The scale is chosen to be $\mu(\mu_F) = M_{JJ}$. The upper set of lines
(blue and pink) are for SP-type pure contact interaction (right most, lower panel).
        }
\label{fig:tot_cs}
\end{figure}
%%%%%%%%%%%

 In figure \ref{fig:tot_cs} we have plotted the total cross section as a function of invariant mass of dijet ($M_{JJ}$). For dijet production, the dijet invariant mass is same as the effective center of mass energy $s$ (due to the {\it ISR} effect). The rapid fall of $\sigma_{SM}$ with $M_{JJ}$ is due to $s^{-1}$ where as the interference cross section ($\sigma_{intf}$) is independent of $s$ (eqn.(\ref{diff_lo_cs_va})) and hence is almost constant. The slow fall of $\sigma_{intf}$ reflects the higher dimensional nature of the contact interaction lagrangian.  Here we have chosen the scale to be $\mu ~(\mu_F) = M_{JJ}$. The pure contact interaction cross section $\sigma_{\eta^2(\xi^2)}$ increases with $M_{JJ}$ due to fact that it is proportional to the $s$ (eqns.(\ref{diff_lo_cs_va},\ref{diff_lo_cs_sp})). The $VA$-type contact interaction cross section increases very slowly with $M_{JJ}$ because of its $V-A$ current structure compare to $SP$-type contact interaction cross section. Consequently larger value of $M_{JJ}$, contact interaction dominates over the SM one. From figure \ref{fig:tot_cs}, it is clear that the cross section is flavor dependent whereas the purely contact cross section  $\sigma_{\eta^2(\xi^2)}$ is flavor independent as we expected. The same argument holds for rest of the analysis.

\begin{figure}[!h]
\centerline{
\epsfxsize=16cm\epsfysize=16.0cm
                     \epsfbox{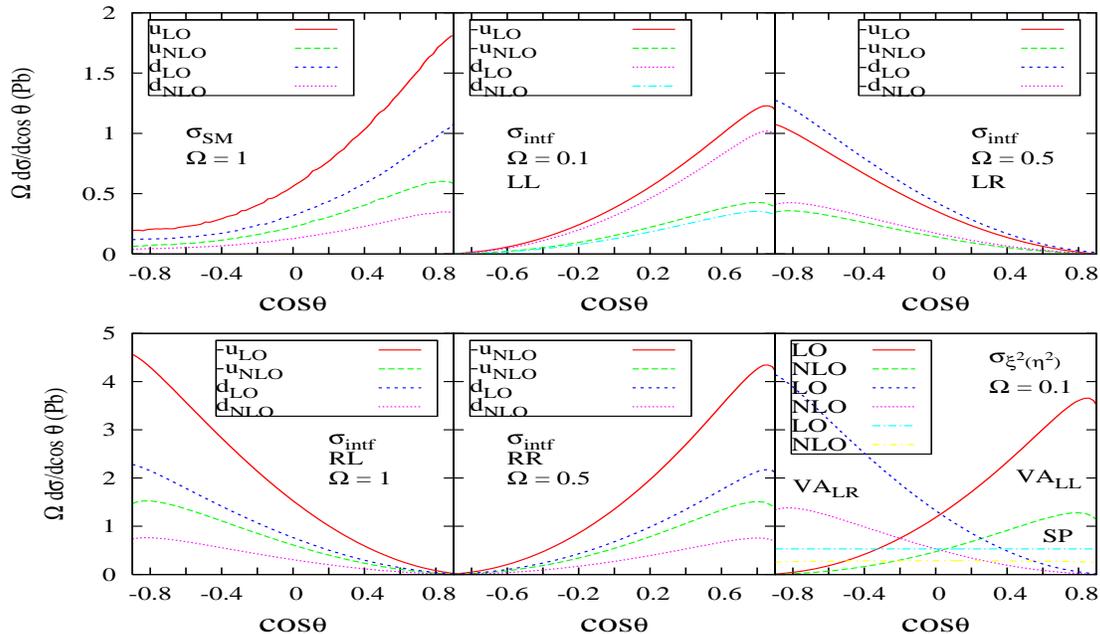}
}
\vspace*{-6.5cm}
\caption{\em The angular distribution for dijet production at $\sqrt{S} = 500$ GeV and $\Lambda = 2$ TeV. The scale $\mu (\mu_F) = P_T$. Here $\Omega$ is a scale factor.
        }
\label{fig:cth_dist}
\end{figure}

\begin{figure}[!h]
\centerline{
\epsfxsize=16cm\epsfysize=16.0cm
                     \epsfbox{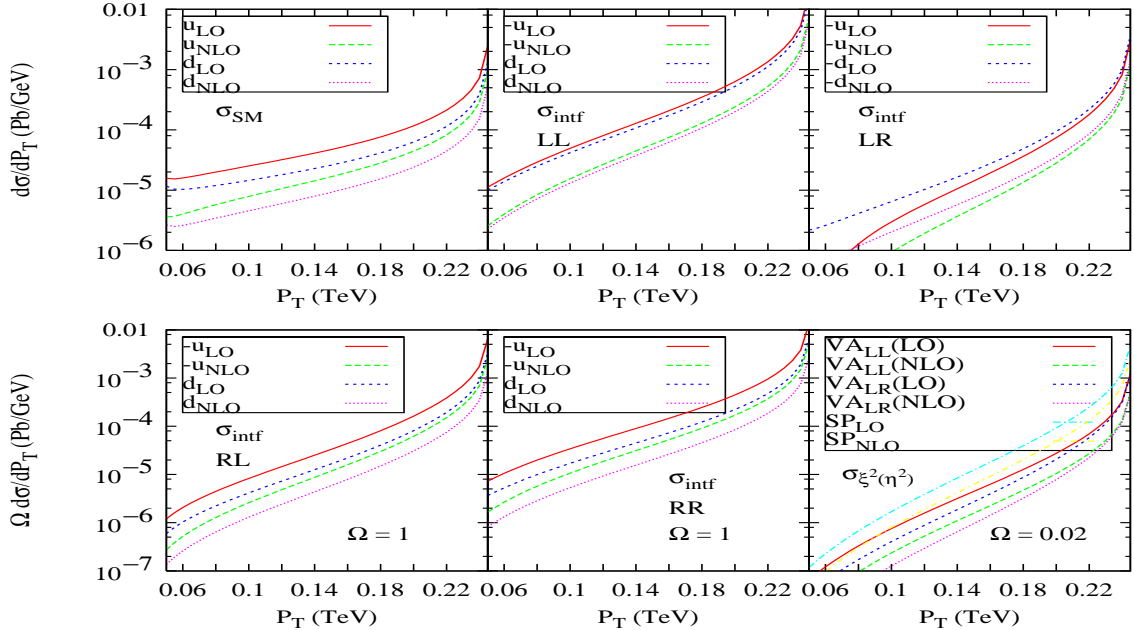}
}
\vspace*{-6.5cm}
\caption{\em Single jet $P_T$ differential distribution for $\sqrt{S} = 500$ GeV, $\Lambda = 2$ TeV and $y_{min} = 0.01$. The scale $\mu (\mu_F) = P_T$. $\Omega$ is a scale factor.
        }
\label{fig:pt_dist}
\end{figure}

      Figure \ref{fig:cth_dist} shows the angular distribution  between beam axis and the jet axis. This distribution  is almost constant for $SP$-type contact interaction because of the fact that the leading order cross section is independent of $\theta$ (which is typical characteristic of scalar vertex) as we see in eqn.(\ref{diff_lo_cs_sp}). This argument holds not only for leading order result but for higher order corrections result as well. The small variation in the NLO result (yellow line) is due to scale variation through the $\alpha_s(\mu^2)$. Whereas this is not so for $VA$-type contact interaction. $VA$-type contact interaction does depend on the $\theta$. These differential distributions for both $VA$-type and $SP$-type contact interaction dominates over the SM piece. For the pure $VA$-type contact interaction, the angular distributions are different for $LL(RR)$ and $LR(RL)$ sector. This is because of the sign of $\cos \theta$ are different(eqn.(\ref{diff_lo_cs_va})). In other words, their chirality structures are quite different. This is also true for $P_T$-distributions (see figure \ref{fig:pt_dist}).
%%%%%%%%%%
%%%%%%%%

In figure \ref{fig:pt_dist} we have shown the transverse momentum ($P_T$)-distribution of a single jet. Since at the leading order, the transverse momentum of the two jets balanced each other ($P_{T_{1}} = P_{T_{2}} = P_T$). For the NLO, the unobserved third parton can be taken infinitely soft for IR safety which is the artifact of fixed order perturbation theory. Therefore this momentum relation still holds even at NLO. From the figure it is clear that as $P_T$ approaches towards $\sqrt{S}/2$ the differential cross section approaches infinity as we expected. Though the interference $P_T$ differential distribution of $RL$ and $LR$ are of the same order or less but $LL$ and $RR$ distributions are of the order one more than the SM whereas for pure contact interaction it is $\sim 10^2$ over the SM for both the cases ($SP$ as well as $VA$).
%%%%%%%

\begin{figure}[!h]
\centerline{
\epsfxsize=16cm\epsfysize=23.0cm
                     \epsfbox{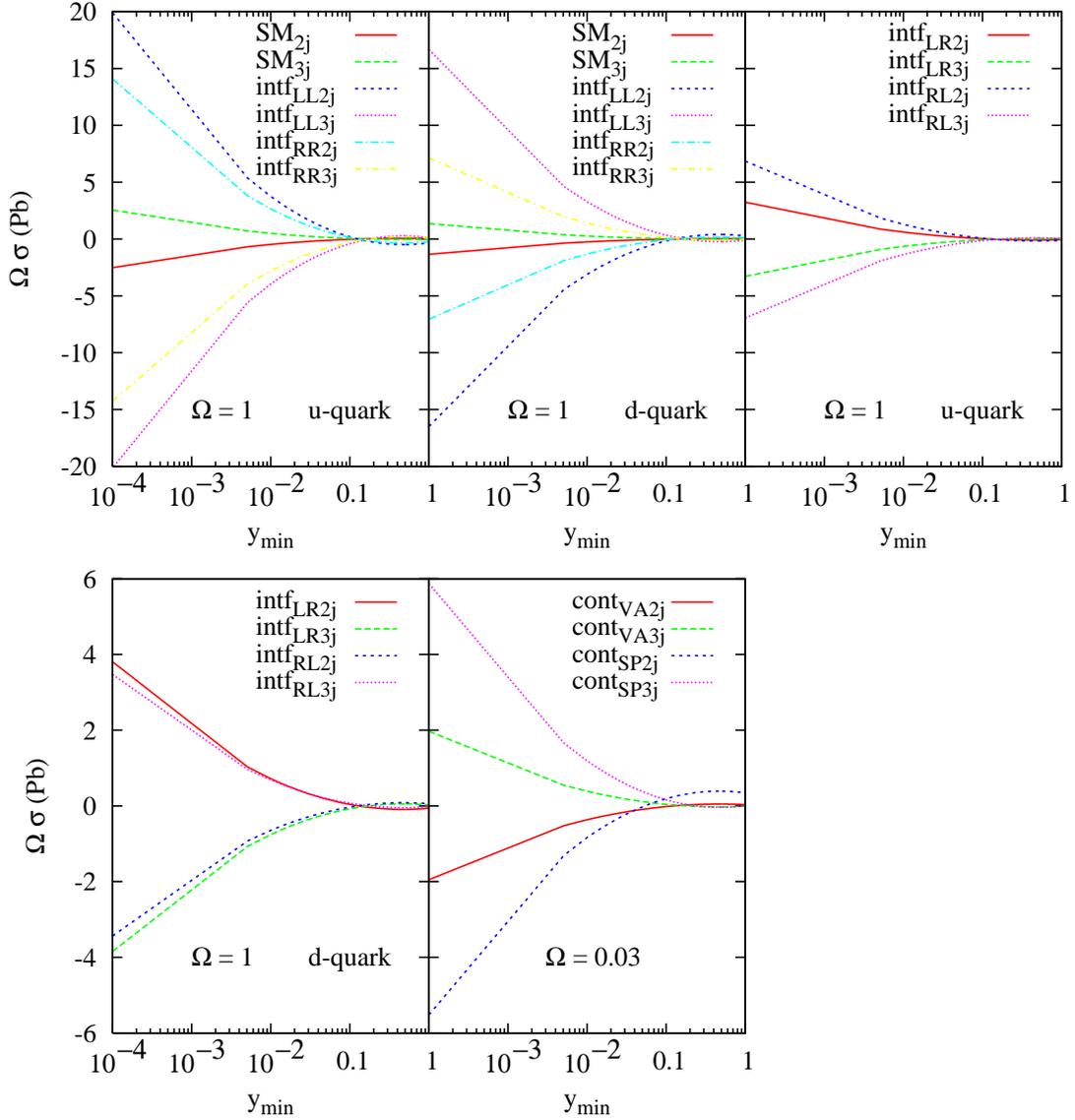}
}
\vspace*{-6.5cm}
\caption{\em Dijet and 3-jet cross section as a function of $y_{min}$ for $\sqrt{s} = 500$ GeV, $\Lambda = 2$ TeV. The scale $\mu (\mu_F) = P_T$. $\Omega$ is a scale factor. Here ``intf'' signifies the interference cross section and ``cont'' signifies the pure contact interaction cross section.
        }
\label{fig:y_dependent}
\end{figure}

In figure \ref{fig:y_dependent} , we have plotted dijet cross section (only ${\cal{O}}(\alpha_s)$) and resolved 3-jet cross section as function of $y_{min}$ (without $ISR$ effect). From the figure, one can easily see that large values of $y_{min}$, both two-parton and three parton cross section produce the meaningful result compared to very small valued $y_{min}$. We have also checked numerically that the sum of these dijet and resolved three jet cross section is independent of $y_{min}$ 
(as it is cleared from the analytic structure in Section \ref{NLO_corr}) which  essentially reproduce the inclusive results for $e^+e^-$ annihilation to quark-antiquark pair upto ${\cal O} (\alpha_s)$.

%%%%%%%%
%%%%%%
%\begin{figure}[!h]
%\centerline{
%\epsfxsize=16cm\epsfysize=16.0cm
%                     \epsfbox{y_dep.ps}
%}
%\vspace*{-2cm}
%\caption{\em K-factor for two jet production as a function of $y_{cut}$.
%        }
%\label{fig:y_dep}
%\end{figure}
%%%

%In figure \ref{fig:y_dep} we have plotted next-to-order $K$-factor as a function $y_{min}$ as defined below:
%\begin{equation}
%K = \sigma_{NLO}(y_{min})/\sigma_{LO}.
%\end{equation}
%From the figure, we see that $K$-factor varies widely for very small values of $y_{min} (< 0.01)$ and almost constant, close to one for $y_{min} > 0.01$. Since $s_{min}$ (or $y_{min}$) is an arbitrary parameter and any physical observable should not depend on it. So in principle one can take $y_{min}$ to be large as close to one. However there is a constraint on $y_{min}$ of the order of $10^{-2}$ for the experimental visibility (detailed discussion can be found \cite{Giele_Glover,kramer_lampe}). For this reason we have chosen $y_{min} = 0.03$ through out our numerical analysis.

\section{Conclusion}
\label{summery}
To summarise, we have performed a systematic calculation of the
next-to-leading order QCD corrections for the dijet production via 
$e^+ e^-$ annihilation in theories with contact
interactions. Contrary to naive expectations, we demonstrate explicitly
that the QCD corrections are meaningful and reliable in the sense that
no undetermined parameters need be introduced.

For the $VA$-type interactions, the analytical structure of the corrections
are similar to those for the SM. However,  a significant dependence
on the flavor structure is found and needs to be carefully
accounted for in obtaining any experimental bounds. For the $SP$-type
interaction, not only are the analytical results  quite different, but 
the results are typically larger than those within the SM. Finally, we have 
investigated the sensitivity of our results to $y_{min}$ (or $s_{min}$ the invariant 
mass cut).

\section*{Acknowledgements}
%DC thanks the
%Department of Science and Technology, India for financial assistance
%under the Swarnajayanti Fellowship grant. 
I thank Prof. Debajyoti Choudhury for suggesting this problem and his meaningful comments.
This work is supported
by an US DOE grant No. DE-FG02-05ER41399 and NATO grant No. PST-CLG. 980342.


\begin{thebibliography}{10}

\bibitem{Kram_Lam_1_6}
W.Bartel {\it et. al.}, Z. Phys. {\bf C33} (1986) 23;\\
S. Bethke {\it et. al.}, Phys.Lett. {\bf B213} (1988) 235;\\
W. Braunschweig {\it et. al.}, Phys.Lett. {\bf B214} (1988) 286;\\
I. Park {\it et. al.}, Phys. Rev. Lett. {\bf 62} (1989) 1713;\\
S. Bethke Z. Phys. {\bf C43} (1989)325 ;\\
S. Komamiya {\it et. al.}, Phys. Rev. Lett. {\bf 64} (1990) 987;\\
M.Z. Akrawy {\it et. al.}, Phys.Lett. {\bf B235} (1990) 289;\\
K. Abe {\it et. al.}, Phys.Lett. {\bf B240} (1990) 232 .

\bibitem{Bel_Cam_Ros_1}
S. Bethke, {\it Proceedings of the High-Energy Physics International Euroconference on Quantam Chromodynamics (QCD 97)}, Montpellier, France, (1997), edted by S. Marison, Nucl. Phys. {\bf B64} (1998) 54;\\
D.Duchesneau, talk given at {\it 29th International conference on High Energy Physics (ICHEP 98)}, Vancouver, Canada, (1998).

\bibitem{susy}
%\cite{Nilles:1983ge}
% \bibitem{Nilles:1983ge}
  H.~P.~Nilles,
  %``Supersymmetry, Supergravity And Particle Physics,''
  Phys.\ Rept.\  {\bf 110} (1984) 1;\\
  %%CITATION = PRPLC,110,1;%%
%\cite{Haber:1984rc}
%\bibitem{Haber:1984rc}
  H.~E.~Haber and G.~L.~Kane,
  %``The Search For Supersymmetry: Probing Physics Beyond The Standard Model,''
  Phys.\ Rept.\  {\bf 117} (1985) 75;\\
  {\em Perspectives in Supersymmetry}, ed. G.L. Kane,
World Scientific (1998); \\
  {\em Theory and Phenomenology of Sparticles}:
M. Drees, R.M. Godbole and P. Roy, World Scientific (2005).
  %%CITATION = PRPLC,117,75;%%

 \bibitem{Pati:1974yy}
  J.~C.~Pati and A.~Salam,
  %``Lepton Number As The Fourth Color,''
  Phys.\ Rev.\ D {\bf 10} (1974) 275.
  %%CITATION = PHRVA,D10,275;%%


\bibitem{GUTS}
%\cite{Georgi:1974sy}
% \bibitem{Georgi:1974sy}
  H.~Georgi and S.~L.~Glashow,
  %``Unity Of All Elementary Particle Forces,''
  Phys.\ Rev.\ Lett.\  {\bf 32} (1974) 438 ;\\
  %%CITATION = PRLTA,32,438;%%
%\cite{Langacker:1980js}
% \bibitem{Langacker:1980js}
  P.~Langacker,
  %``Grand Unified Theories And Proton Decay,''
  Phys.\ Rept.\  {\bf 72} (1981) 185.
  %%CITATION = PRPLC,72,185;%%


\bibitem{preon}
H. Harari and N. Seiberg, Phys.Lett. {\bf B98} (1981) 269;\\
M.E.  Peskin, {\it in proceedings of the 1981 International Symposium on Lepton and Photon Interaction at High Energy,}
W.Pfeil, ed., p880 (Bonn, 1981);\\
L. Lyons, Oxford University Publication 52/82 (June 1982).

\bibitem{preon_binding}
G. 't Hooft, in Recent Developements in Gauge Theories;\\
G. 't Hooft {\it et al.}, ads. (Plenum Press, New York,1980).

\bibitem{comp_mod}
Jogesh C. Pati, Abdus Salam and J.A. Strathdee Phys.Lett. {\bf B59} (1975) 265;\\
H. Fritzsch and G. Mandelbaum, Phys.Lett. {\bf B102} (1981) 319;\\
W. Buchmuller, R.D. Peccei and T. Yanagida, Phys.Lett. {\bf B124} (1983) 67;
Nucl.Phys.{\bf B227} (1983) 503; Nucl.Phys.{\bf B237} (1984) 53;\\
 U. Baur and H. Fritzsch, Phys.Lett. {\bf B134} (1984) 105;\\
Xiaoyuan Li and R.E. Marshak, Nucl.Phys.{\bf B268} (1986) 383;\\
 I. Bars, J.F. Gunion and M. Kwan Nucl.Phys.{\bf B269} (1986) 421;\\
 G. Domokos and S. Kovesi-Domokos, Phys.Lett.{\bf B266} (1991) 87;\\
Jonathan L. Rosner and Davison E. Soper Phys.Rev.{\bf D45} (1992) 3206;\\
Markus A. Luty and Rabindra N. Mohapatra, Phys.Lett.{\bf B396} (1997) 161 [hep-ph/9611343];\\
K. Hagiwara, K. Hikasa and M. Tanabashi, Phys.Rev.{\bf D66} (2002) 010001; Phys.Lett.{\bf B592} (2004) 1.

\bibitem{Additional_comp}
For a review and additional references, see R.R. Volkas and G.C. Joshi, Phys. Rep.
{\bf 159} (1988) 303.


\bibitem{cont_inter}
R. Ruckle, Phys. Lett. {\bf B129} (1983) 363; Nucl. Phys. {\bf B234} (1984) 91;\\
W. Buchmuller, R. Ruckle and D. Wyler, Phys. Lett. {\bf B191} (1987) 442;\\
P. Haberl, F. Schrempp and H. U. Martyn, in {\it Physics at HERA}, eds.
W. Buchmuller and G. Ingelman, DESY (1991) P.1133;\\
W. Buchmuller and D. Wyler, Phys. Lett. {\bf B407} (1997) 147  
[hep-ph/970431];\\
N.G.Deshpande, Bhaskar Dutta and 
Xiao-Gang He, Phys. Lett. {\bf B408} (1997) 288 
[hep-ph/9705236].

\bibitem{Taekoon}
T. Lee, Phys. Rev. D {\bf 55} (1997) 2591 [hep-ph/9605429].

\bibitem{llmodel}
E. Eichten, K.D. Lane and M.E. Peskin, Phys. Rev. Lett. {\bf 50} (1983) 811 ;\\
E. Eichten, I. Hinchliffe, K.D. Lane and C. Quigg, Rev. Mod. Phys.
{\bf 56} (1984) 579. 

\bibitem{leptoquarks}
% \bibitem{Buchmuller:1986iq}
  W.~Buchmuller and D.~Wyler,
  %``Constraints On SU(5) Type Leptoquarks,''
  Phys.\ Lett.\ B {\bf 177} (1986) 377;\\
  %%CITATION = PHLTA,B177,377;%%
% \bibitem{Buchmuller:1986zs}
  W.~Buchmuller, R.~Ruckl and D.~Wyler,
  %``Leptoquarks In Lepton Quark Collisions,''
  Phys.\ Lett.\ B {\bf 191} (1987) 442;
  [Erratum-ibid.\ B {\bf 448} (1999) 320 ];\\
  %%CITATION = PHLTA,B191,442;%%
%\cite{Hewett:1997ce}
% \bibitem{Hewett:1997ce}
  J.~L.~Hewett and T.~G.~Rizzo,
  %``Much ado about leptoquarks: A comprehensive analysis,''
  Phys.\ Rev.\ D {\bf 56} (1997) 5709
  [hep-ph/9703337].
  %%CITATION = HEP-PH 9703337;%%

\bibitem{fayet} 
% \bibitem{Fayet:1977yc}
  P.~Fayet,
  %``Spontaneously Broken Supersymmetric Theories Of Weak, Electromagnetic And
  %Strong Interactions,''
  Phys.\ Lett.\ B {\bf 69}  (1977) 489;\\
  %%CITATION = PHLTA,B69,489;%%
%\cite{Farrar:1978xj}
% \bibitem{Farrar:1978xj}
  G.~R.~Farrar and P.~Fayet,
  %``Phenomenology Of The Production, Decay, And Detection Of New Hadronic
  %States Associated With Supersymmetry,''
  Phys.\ Lett.\ B {\bf 76} (1978) 575.
  %%CITATION = PHLTA,B76,575;%%

\bibitem{Hasenfratz:1987tk}
  P.~Hasenfratz and J.~Nager,
  %``The Cutoff Dependence Of The Higgs Meson Mass And The Onset Of New Physics
  %In The Standard Model,''
  Z.\ Phys.\ C {\bf 37} (1988) 477.
  %%CITATION = ZEPYA,C37,477;%%

\bibitem{HERA2}
  H1 Collaboration, C. Adloff {\it et al.}, Z. Phys. C{\bf 74} (1997) 191 [hep-ex/9702012];\\
  ZEUS Collaboration, J. Breitweg {\it et al.}, Z. Phys. C{\bf 74} (1997) 207 [hep-ex/9702015].

\bibitem{opal}
CELLO Collaboration, H. J.Behrend {\it et al.}, Phys. Lett. {\bf B191} (1987) 209;Phys. Lett. {\bf B222} (1989) 163;Z. Phys. {\bf C51} (1991) 143;Z. Phys. {\bf C51} (1991) 149;\\
VENUS Collaboration, K. Abe {\it et al.}, Phys. Lett. {\bf B232} (1989) 425;\\
TOPAZ Collaboration, I. Adachi {\it et al.}, Phys. Lett. {\bf B255} (1991) 613;\\
OPAL Collaboration,  G. Alexander {\it et al.}, Phys. Lett. {\bf B387} (1996) 432.

\bibitem{opal2}
OPAL Collaboration,  K. Ackerstaff {\it et al.}, Phys. Lett. {\bf B391} (1997) 221. 

\bibitem{cdfprl}
CDF Collaboration (F. Abe {\it et al.}), Phys. Rev. {\bf D44} (1991) 29;
{\it ibid.} {\bf 68} (1992) 1463;\\
Phys. Rev. Lett. {\bf 69} (1992) 28;
ibid. {\bf 79} (1997) 2198.


\bibitem{Drell_Yan}
S.D. Drell and T.M. Yan, Phys. Rev. Lett. {\bf 25} (1970) 316;\\
J.H. Christenson {\it et al.}, {\it ibid.} {\bf 25} (1970) 1523;\\
L.M. Lederman and B.G. Pope, {\it ibid.} {\bf 27} (1971) 765.

\bibitem{D0_coll}
D0 Collaboration Phys. Rev. Lett. {\bf 82} (1999) 4769 [hep-ex/9812010].

\bibitem{NKMondal}
S. Jain, A.K. Gupta and N.K. Mondal, Phys. Rev. D {\bf 62} (2000) 095003
[hep-ex/0005025].

\bibitem{3rd_paper} D. Choudhury, S. Majhi and V. Ravindran, JHEP {\bf 0601:027},2006
[hep-ph/0509057]

\bibitem{Giele_Glover}
W.T. Giele and E.W.N. Glover, Phys. Rev. {\bf D46} (1992) 1980.

\bibitem{sterman}
G. Sterman and S. Weinberg, Phys. Rev. Lett. {\bf 39} (1977) 1436.

\bibitem{kramer_lampe}
G. Kramer and B. Lampe, Z. Phys. {\bf C34} (1987) 497 and reference therein.


%\bibitem{ravi_gravi}
%G. Altarelli, R.K.Ellis and G. Martinelli,
%Nucl. Phys. {\bf B157} (1979) 461;\\
%B.Humpert and W.L. van Neerven, Phys. Lett. {\bf B84} (1979) 327;
%[Errat. {\bf B85} (1979) 471]; ibid. {\bf B89} (1979) 69; Nucl. Phys. {\bf B
%184} (1981) 225;\\
%J.Kubar, M. le Bellac, J.L.Meunier and G. Plaut, Nucl. Phys. {\bf B175}
%(1980) 251;\\
%P. Aurenche and P. Chiapetta, Z.Phys. {\bf C34} (1987) 201;\\
%P.J.Sutton, A.D.Martin, R.G. Roberts W.J.Stirling, Phys. Rev. {\bf D45}
%(1992) 2349;\\
%P.J. Rijken and W.L. van Neerven, Phys. Rev. {\bf D51} (1995) 44
%[hep-ph/9408366];\\
% P. Mathews, V. Ravindran, K. Sridhar and W.L. van Neerven,
%Nucl. Phys. {\bf B716} (2005) 128 [hep-ph/0411018].
%
%\bibitem{Altareli_parisi}
%G. Altarelli and G. Parisi, Nucl. Phys. {\bf B126} (1977) 298.

\bibitem{1st_paper}
D. Choudhury, S.Majhi and V. Ravindran, Nucl. Phys. {\bf B660}
(2003) 343 [hep-ph/0207247];\\
Li Lin Yang, Chong Sheng Li, Jian Jun Liu, Qiang Li, Phys.Rev.{\bf D72} (2005) 074026
[hep-ph/0507331].

%\bibitem{Giele_Glover}
%W.T. Giele and E.W.N. Glover, Phys. Rev. {\bf D46} (1992) 1980.

%\bibitem{sterman}
%G. Sterman and S. Weinberg, Phys. Rev. Lett. {\bf 39} (1977) 1436.

%\bibitem{kramer_lampe}
%G. Kramer and B. Lampe, Z. Phys. {\bf C34} (1987) 497.

\bibitem{ravi_gravi}
G. Altarelli, R.K.Ellis and G. Martinelli,
Nucl. Phys. {\bf B157} (1979) 461;\\
B.Humpert and W.L. van Neerven, Phys. Lett. {\bf B84} (1979) 327;
[Errat. {\bf B85} (1979) 471]; ibid. {\bf B89} (1979) 69; Nucl. Phys. {\bf B
184} (1981) 225;\\
J.Kubar, M. le Bellac, J.L.Meunier and G. Plaut, Nucl. Phys. {\bf B175}
(1980) 251;\\
P. Aurenche and P. Chiapetta, Z.Phys. {\bf C34} (1987) 201;\\
P.J.Sutton, A.D.Martin, R.G. Roberts W.J.Stirling, Phys. Rev. {\bf D45}
(1992) 2349;\\
P.J. Rijken and W.L. van Neerven, Phys. Rev. {\bf D51} (1995) 44
[hep-ph/9408366];\\
 P. Mathews, V. Ravindran, K. Sridhar and W.L. van Neerven,
Nucl. Phys. {\bf B716} (2005) 128 [hep-ph/0411018].
%

\bibitem{catani} S. Catani, [hep-ph/9411361].

\bibitem{g_kramer}
G. Kramer, Springer Tracks in Moder Physics, Vol.102, Berlin 1984.

\bibitem{ISR} E.A. Kuraev and V.S. Fadin, Sov. J. Nucl. Phys. {\bf 41} (1985) 466 
[Yad.Fiz. {\bf 41} (1986) 733] ;
O. Nicrosini and L. Trentadue, Phys. Lett. {\bf B196} (1987) 551.

\end{thebibliography}
\end{document}